\documentclass[aps,superscriptaddress,twocolumn,showpacs,preprintnumbers,amsmath,amssymb]{revtex4}
\usepackage{pst-all} 
\usepackage{multido}
\usepackage[dvips]{graphicx}
\usepackage{eucal}
\usepackage{bm}
\usepackage{float}
\usepackage{color}

\newcommand{\comment}[1]{}

\begin{document}
\title{
Eccentricity fluctuations from the Color Glass Condensate at RHIC and LHC
}

\medskip

\author{Hans-Joachim Drescher}
\affiliation{
Frankfurt Institute for Advanced Studies (FIAS),
Johann Wolfgang Goethe-Universit\"at,
Max-von-Laue-Str.~1, 60438  Frankfurt am Main, Germany
}
\author{Yasushi Nara}
\affiliation{
Akita International University
193-2 Okutsubakidai, Yuwa-Tsubakigawa,
Akita-city, Akita 010-1211 Japan
}

\begin{abstract}
In this brief note, we determine the fluctuations of the initial
eccentricity in heavy-ion collisions caused by fluctuations of the
nucleon configurations. This is done via a Monte-Carlo implementation
of a Color Glass Condensate $k_t$-factorization approach. The
eccentricity fluctuations are found to nearly saturate elliptic flow
fluctuations measured recently at RHIC. Extrapolations to LHC energies
are shown.
\end{abstract}

\pacs{12.38.Mh,24.85.+p,25.75.Ld,25.75.-q}

\maketitle
\section{Introduction}

In a high energy non-central heavy ion collision the asymmetry of the
coordinate space, the overlap area, is transferred into an asymmetry
in momentum space, and measured as the elliptic flow
$v_2=\langle\cos(2\phi)\rangle$.  The initial asymmetry in coordinate
space is characterized by the eccentricity,
\begin{equation}
\label{eq:ecc}
\varepsilon =
\frac{\langle r_y^{\,2}{-}r_x^{\,2}\rangle}
     {\langle r_y^{\,2}{+}r_x^{\,2}\rangle}~.
\end{equation}
where the brackets $\langle ...\rangle$ indicate an average over the
transverse plane, using some appropriate
weight. Here, we use the number density of produced gluons.

In ideal hydrodynamics, assuming a short thermalization time, the
final elliptic flow is proportional to the initial eccentricity $v_2 =
c ~ \varepsilon$. The proportionality constant depends on the equation
of state but is roughly $c=0.2$~\cite{Ollitrault:1992bk}.

Fluctuations of the eccentricity therefore should translate into
fluctuations of the elliptic flow~\cite{Socolowski:2004hw}. Recently,
these $v_2$ fluctuations have been measured by the PHOBOS and the STAR
collaborations \cite{Alver:2007qw,Sorensen:2006nw}.

In this note, we examine the fluctuations of $\varepsilon$ based on
the Monte Carlo KLN model introduced in Ref.\ \cite{Drescher:2006ca} and
compare to standard Glauber-model results (see, for
example~\cite{Bhalerao:2006tp,Broniowski:2007ft}).

\section{Improvements in the MC-KLN model}
In Ref. \cite{Drescher:2006ca} we introduced a Monte Carlo
implementation of the Kharzeev-Levin-Nardi
(MC-KLN)~\cite{Kharzeev:2004if} approach to particle production in
heavy ion collisions. Gluon production is calculated individually for
each configuration of nucleons in the colliding nuclei. Thanks to the
implementation of perturbative gluon saturation in this approach, the
multiplicity can be determined via the well-known $k_t$-factorization
formula~\cite{Kharzeev:2004if} without the need to introduce infrared
cutoffs (and additional models for the soft regime). The saturation
scale is taken to be proportional to the local density of nucleons
which, in turn, is measured by counting nucleons in a given sampling
area. However, if the radius of the sampling area is $r_{\rm
  max}=\sqrt{\sigma_{\mathrm inel}/\pi}$, one overestimates the
interaction probability especially in the periphery, since nucleon
pairs can have a distance up to $2r_{\rm max}$. Therefore, we improved
on our previous model by rejecting those pairs with $r>r_{\rm
  max}$. In the $p+p$ limit this results in an additional factor
$0.58$ which is very close to the value found in
Refs. \cite{Broniowski:2007ft,Kharzeev:2004if} by accounting for the
difference between the inelastic and the geometric cross section of a
nucleon.  We further assume here that $\sigma_{\mathrm inel}=42$~mb at
full RHIC energy ($\sqrt{s_{NN}}=200$~GeV), and $\sigma_{\mathrm
  inel}=66$~mb at LHC energy ($\sqrt{s_{NN}}=5500$~GeV).

\begin{figure}[tb]
\begin{center}
\includegraphics[width=8cm,clip]{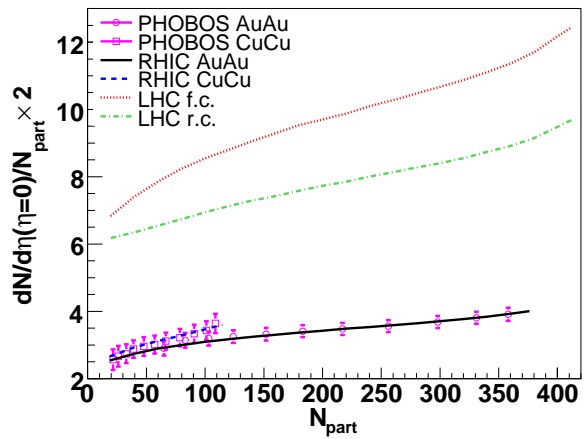}
\caption{(Color online) Multiplicity for Au+Au and Cu+Cu collisions at
  200 GeV and PbPb collisions at 5500 GeV. The data is from the PHOBOS collaboration\cite{Back:2002uc,Roland:2005ei}. }
\label{fig:mult}
\end{center}
\end{figure} 
This refined treatment allows for an excellent description of the
charged multiplicity at RHIC over the entire range of centralities
(for both Cu and Au nuclei), essentially down to $p+p$
collisions. Fig.~\ref{fig:mult} depicts our results for full RHIC
energy, as well as an extrapolation to Pb+Pb collisions at LHC
energy. Since there is some uncertainty regarding the evolution of the
saturation scale, we show results for both fixed coupling evolution,
$Q_s^2=Q_{s,0}^2(x_0/x)^\lambda$ with $\lambda=0.28$, and running
coupling evolution of $Q_s^2$ (see e.g.~\cite{Kharzeev:2004if}). For
the latter case, the initial condition $Q_{s,0}$ and $x_0$ was set
such that at RHIC energy $Q_s$ agrees with previous estimates.

\section{Fluctuations of the initial eccentricity}

The participant eccentricity $\varepsilon_{\mathrm part}$, which
corrects for fluctuations of the major axes and of the center of mass of
the overlap region, is defined by:
\begin{equation}
\varepsilon_\mathrm{part}
= \frac{\sqrt{(\sigma^2_y - \sigma^2_x)^2+4\sigma^2_{xy}}}
       {\sigma^x_y+\sigma^2_x},
\end{equation}
The fluctuations of this variable for a given centrality class (here
defined by the number of participants) are determined via
\begin{equation}
  \sigma_{\varepsilon_\mathrm{part}}=\sqrt{\langle
  \varepsilon_\mathrm{part}^2 \rangle 
-
\langle
  \varepsilon_\mathrm{part} \rangle^2 }~.
\end{equation}

\begin{figure}[t]
\begin{center}
\includegraphics[width=8cm,clip]{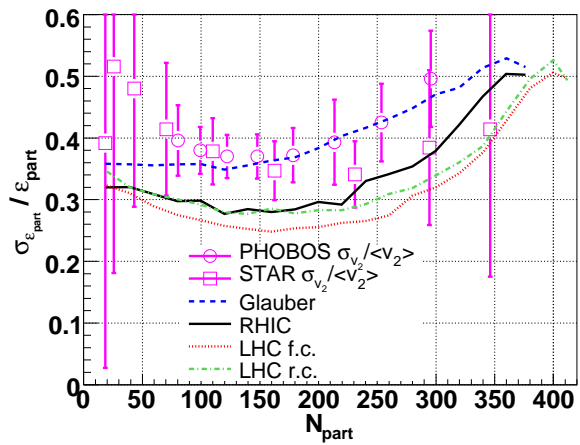}
\caption{(Color online)
Relative fluctuations of the eccentricity as a function of centrality
in Au+Au / Pb+Pb collisions.
}\label{fig:fluc}
\end{center}
\end{figure}

Fig. \ref{fig:fluc} shows the result together with data from PHOBOS
\cite{Alver:2007qw} and STAR \cite{Sorensen:2006nw}, and a simple
Glauber model, where the number density of gluons scales with the
number of participants $N_{\mathrm part}$ (note that this model fails
to account for the growth of $dN/d\eta/N_{\mathrm part}$ with
centrality seen in Fig.~\ref{fig:mult}). These measurements are rather
difficult, and therefore the error bars are quite large, as is the
discrepancy between experiments, especially at high centralities
where neither the Glauber model, nor the CGC result can be ruled
out. For semi-central collisions, the CGC predicts somewhat lower
relative fluctuations than the Glauber model. We note that
$\sigma_{\varepsilon_\mathrm{part}}$ itself is quite independent of
the underlying model and energy. The main reason for the lower {\em
relative} eccentricity fluctuations in the MC-KLN model is the larger
average eccentricity for semi-central Au+Au collisions in this
approach, see the discussion in
refs.~\cite{Drescher:2006ca,Drescher:2006pi}.

To check for other possible sources of fluctuations in the participant
eccentricity, we implemented additional Poissonian (uncorrelated)
fluctuations of the number of gluons produced at a given point in the
transverse plane. These may arise, for example, from fluctuations of
the gluon evolution ladders. However, we found that they did not
noticeably affect $\sigma_{\varepsilon_\mathrm{part}}$. One should
also keep in mind that so-called non-flow effects may increase
fluctuations of the measured $v_2$. Moreover, hydrodynamic
fluctuations may contribute to $\sigma_{v_2}$ as well~\cite{Vogel:2007yq}.
Hence, $\sigma_{v_2}/v_2$ should be viewed only as an upper limit for
$\sigma_{\varepsilon_\mathrm{part}}/\varepsilon_\mathrm{part}$.

\section{Summary}

We have calculated the fluctuations of the initial eccentricity within
a simple Glauber model and for a Color Glass Condensate approach which
includes fluctuations in the positions of the hard sources (nucleons).
Both models predict eccentricity fluctuations which nearly
saturate the experimentally measured fluctuations of the elliptic flow.
The CGC approach gives slightly lower {\em relative} fluctuations than
the Glauber model, which is largely due to a higher average
eccentricity $\varepsilon_\mathrm{part}$. Their magnitude at LHC
energy is similar.

\begin{acknowledgments}
The authors are thankful to Adrian Dumitru for useful comments. 
HJD acknowledges support from BMBF grant 05 CU5RI1/3.
\end{acknowledgments}

\end{document}